\documentclass[11pt,twoside]{article}
\usepackage{graphicx,epsfig,natbib}
\usepackage{CS18}

\newcommand{\COBOLD}{{\tt CO$^5$BOLD}}

\newcommand{\LHD}{{\tt LHD}}
\newcommand{\MARCS}{{\tt MARCS}}

\newcommand{\ATLAS}{{\tt ATLAS9}}

\newcommand{\teff}{\ensuremath{T_{\mathrm{eff}}}}
\newcommand{\logg}{\ensuremath{\mathrm{log}\,g}}
\newcommand{\MoH}{\ensuremath{\left[\mathrm{M}/\mathrm{H}\right]}}
\newcommand{\FeH}{\ensuremath{\left[\mathrm{Fe}/\mathrm{H}\right]}}
\newcommand{\OFe}{\ensuremath{\left[\mathrm{O}/\mathrm{Fe}\right]}}
\newcommand{\xtmean}[1]{\ensuremath{\left\langle #1\right\rangle}}

\begin{document}

%

\title{Oxygen in the early Galaxy: OH lines as tracers of oxygen abundance
in extremely metal-poor giant stars}

\author{A~.Ku\v{c}inskas$^{1}$, V.~Dobrovolskas$^{1}$, P.~Bonifacio$^{2}$, E.~Caffau$^2$, H.-G.~Ludwig$^3$, M.~Steffen$^4$, M.~Spite$^2$}

\affil{$^1$Institute of Theoretical Physics and Astronomy, Vilnius University, A.~Go\v{s}tauto 12, Vilnius LT-01108, Lithuania}
\affil{$^2$GEPI, Observatoire de Paris, CNRS, Universit\'{e} Paris Diderot, Place Jules Janssen, 92190 Meudon, France}
\affil{$^3$Landessternwarte -- Zentrum f\"ur Astronomie der Universit\"at Heidelberg, K\"{o}nigstuhl 12, D-69117 Heidelberg, Germany}
\affil{$^4$Leibniz-Institut f\"ur Astrophysik Potsdam, An der Sternwarte 16, D-14482 Potsdam, Germany}

\begin{abstract}
Oxygen is a powerful tracer element of Galactic chemical evolution. Unfortunately, only a few oxygen lines are available in the ultraviolet-infrared stellar spectra for the reliable determination of its abundance. Moreover, oxygen abundances obtained using different spectral lines often disagree significantly. In this contribution we therefore investigate whether the inadequate treatment of convection in 1D hydrostatic model atmospheres used in the abundance determinations may be responsible for this disagreement. For this purpose, we used VLT~CRIRES spectra of three EMP giants, as well as 3D hydrodynamical \COBOLD\ and 1D hydrostatic \LHD\ model atmospheres, to investigate the role of convection in the formation of infrared (IR) OH lines. Our results show that the presence of convection leads to significantly stronger IR~OH lines. As a result, the difference in the oxygen abundance determined from IR~OH lines with 3D hydrodynamical and classical 1D hydrostatic model atmospheres may reach $-0.2\,\dots\,-0.3$~dex. In case of the three EMP giants studied here, we obtain a good agrement between the 3D~LTE oxygen abundances determined by us using vibrational-rotational IR~OH lines in the spectral range of 1514--1626~nm, and oxygen abundances determined from forbidden [O~I] 630~nm line in previous studies.
\end{abstract}

\section{Introduction}

Although oxygen is an important tracer of Galactic chemical evolution, only a few spectral lines are available for its diagnostics in extremely metal-poor (EMP) stars. In red giant stars, [O~I] 630~nm line can be used, however, in the EMP regime it is very weak and frequently blended with telluric oxygen lines ([O~I] and ${\rm O}_2$ band). Alternatively, oxygen abundances can be derived from ultraviolet (UV) and infrared (IR) OH lines but results obtained using molecular and atomic lines are frequently discrepant: for example, abundances obtained from IR~OH lines are typically $\sim0.3-0.4$~dex higher than those determined from the forbidden [O~I] 630~nm line.

One possible reason behind these discrepancies is that 1D hydrostatic model atmospheres that are used in the abundance analysis do not account for the horizontal fluctuations of thermodynamic quantities arising from convective motions in stellar atmospheres. Since formation of OH lines is very sensitive to the effects of convection \citep[see, e.g.,][]{CAT07,DKS13}, one may expect that application of 3D hydrodynamical model atmospheres in the abundance analysis may diminish, or even eliminate, the differences in oxygen abundances obtained from atomic and molecular lines.

In this contribution we therefore study whether a more realistic modeling of convection may bring the oxygen abundances obtained from molecular IR~OH lines into better agreement with those determined using forbidden [O~I] 630~nm line (note that [O~I] 630~nm line is insensitive to 3D hydrodynamical \textit{and} NLTE effects, see, e.g., \citealt[][]{A05}). To this end, we used spectra of three EMP giants obtained with the VLT~CRIRES spectrograph, as well as 3D hydrodynamical \COBOLD\ and 1D hydrostatic \LHD\ model atmospheres, to better understand the role of convection in the formation of IR~OH lines. All results obtained in this analysis will be presented in a forthcoming paper \citep[][]{DKB14}, while in this contribution we briefly summarize some of the most important findings.

\section{Methodology}

\begin{figure}
\centering
\includegraphics[width=12cm]{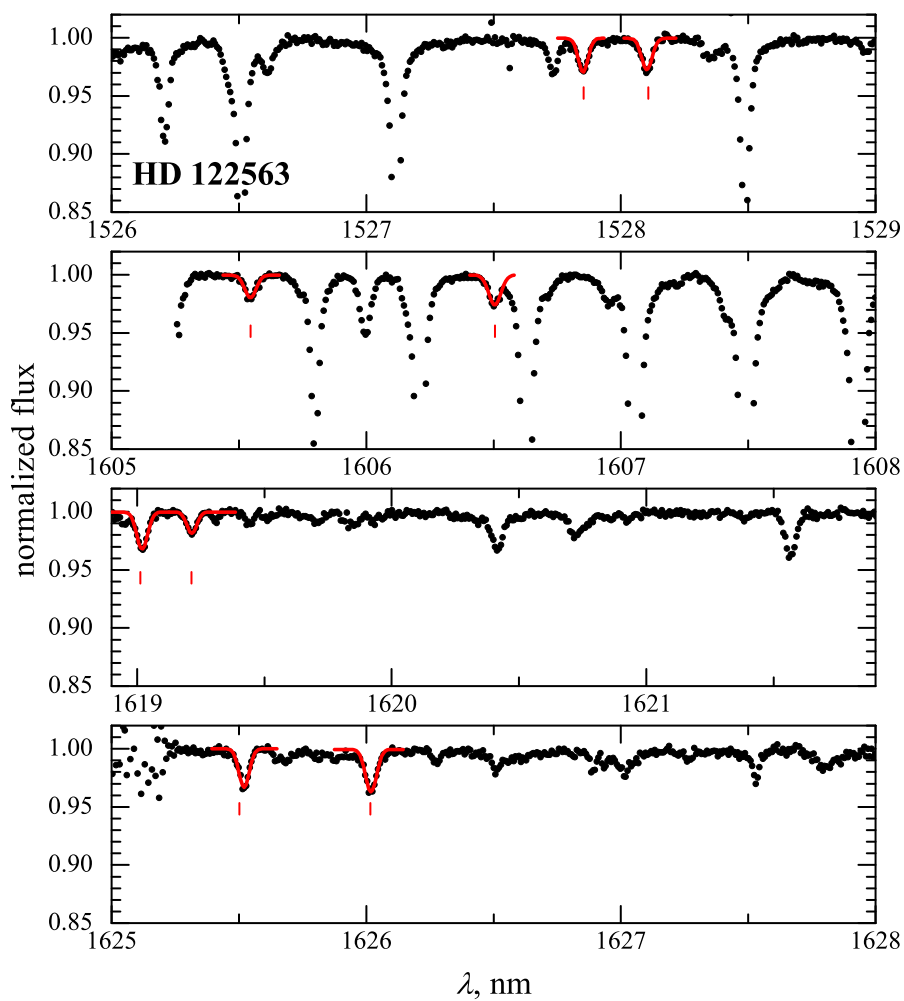}
\caption{Observed VLT~CRIRES spectrum of HD~122563 (black dots). Gaussian profiles fits to IR~OH lines used in the oxygen abundance determination are shown in red.\label{fig:spectrum}}
\end{figure}

We used high resolution ($R = 50\,000$) and high signal-to-noise ratio ($S/N\approx400-600$) near-infrared ($H$-band) spectra of 3 EMP stars ($\FeH=-2.6$ to --3.1, see Table~1) obtained with the CRIRES spectrograph at the VLT~UT1 telescope (program ID 089.D-0079(A); a sample spectrum is shown in Fig.~\ref{fig:spectrum}). For each star, oxygen abundances were determined using the measured equivalent widths of vibrational-rotational IR~OH lines ($X^{-1}\Pi$) from the first-overtone sequence located in the spectral range of $1527-1626$~nm. Atmospheric parameters and iron abundances used in the abundance determination were taken from the literature.

Abundance analysis was done in 1D~LTE with the \ATLAS\ model atmospheres. The obtained 1D~LTE oxygen abundances were corrected for the 3D hydrodynamical effects, by applying 3D--1D abundance corrections to oxygen abundances determined from each individual IR~OH line. The 3D--1D abundance corrections (i.e., differences in the abundance of a given chemical element obtained from the same spectral line using 3D hydrodynamical and 1D hydrostatic model atmospheres) were computed with the 3D hydrodynamical \COBOLD\ \citep{FSL12} and 1D hydrostatic \LHD\ model atmosphere codes \citep{CLS08}. The two types of model atmospheres shared identical elemental abundances, atmospheric parameters, equation of state, and opacities. Abundance corrections were derived using a single 3D hydrodynamical model atmosphere computed using the following atmospheric parameters: $\teff=4600$~K, $\log g=1.6$, $\MoH=-2.5$. We note that differences between the atmospheric parameters of this 3D hydrodynamical model and those of the target EMP stars studied here did not exceed $\Delta \teff=150$~K, $\Delta \log g = 0.2$, and $\Delta \MoH = 0.6$. Such differences would lead to only minor deviations in the abundance corrections, the size of which (typically, a few hundredths of a dex) would render them unimportant in the context of the present study. The 3D hydrodynamical \COBOLD\ model atmosphere used in our work had a spatial resolution of $220\times220\times280$ elements, corresponding to the physical size of $3.85\times3.85\times2.21$~Gm in $x,y,z$ directions, respectively. The solution of radiative transfer equation was made using opacity binning technique \citep[][]{N82,L92}, by sorting opacities into 6 bins. The opacities were taken from the \MARCS\ model atmosphere package \citep[][]{GEK08}.

A sample VLT~CRIRES spectrum with the Gaussian fits to IR~OH lines is shown in Fig.~\ref{fig:spectrum}. The obtained 1D~LTE and 3D~LTE oxygen abundances, as well as abundances determined using forbidden [O~I] 630~nm line (literature data), are provided in Table~\ref{table:stars-abundances}.

\begin{table}[tb]
\caption{List of observed EMP stars and determined oxygen abundances.}
\label{table:stars-abundances}
\smallskip
\begin{center}
\scalebox{0.8}{
\setlength{\tabcolsep}{4pt}
\begin{tabular}{lcccccccc}
\hline
\noalign{\smallskip}
Star        &  $V$  &  $S/N$  & \teff, K  & \logg  & \FeH   & \multicolumn{2}{c}{Abundances from OH lines}           & Abundances from     \\
            &     &         &           &       & 1D LTE  & \multicolumn{2}{c}{(this work)}                          & [O I] 630 nm line   \\
            &     &         &           &       &         & $\OFe^{\rm{1D\,LTE}}$    &   $\OFe^{\rm{3D\,LTE}}$       & $\OFe^{\rm{1D\,LTE}}$ \\
\noalign{\smallskip}
\hline
\noalign{\smallskip}
HD 122563   & 6.2 & 560    & $4600^a$   & $1.6^a$  & $-2.8^b$  & $0.79 \pm 0.12$  & $0.56 \pm 0.13$ & $0.53^c$; $0.60^d$; $0.62^e$   \\
HD 186478   & 9.2 & 440    & $4700^b$   & $1.3^b$  & $-2.6^b$  & $1.02 \pm 0.10$  & $0.76 \pm 0.09$ & $0.75 \pm 0.11^e$              \\
BD -18:5550 & 9.4 & 600    & $4750^b$   & $1.4^b$  & $-3.1^b$  & $1.07 \pm 0.12$  & $0.84 \pm 0.15$ & $0.42 \pm 0.24^e$              \\
\noalign{\smallskip}
\hline
\end{tabular}
}
\end{center}
\begin{list}{}{}
\item[$^{\mathrm{a}}$] \citet{CTB12}; $^{\mathrm{b}}$ \citet{CDS04}; $^{\mathrm{c}}$ \citet{SKP91}; $^{\mathrm{d}}$ \citet{BMS03}; $^{\mathrm{e}}$ \citet{SCP05}.
\end{list}
\end{table}

\begin{figure}
\centering
\includegraphics[width=12cm]{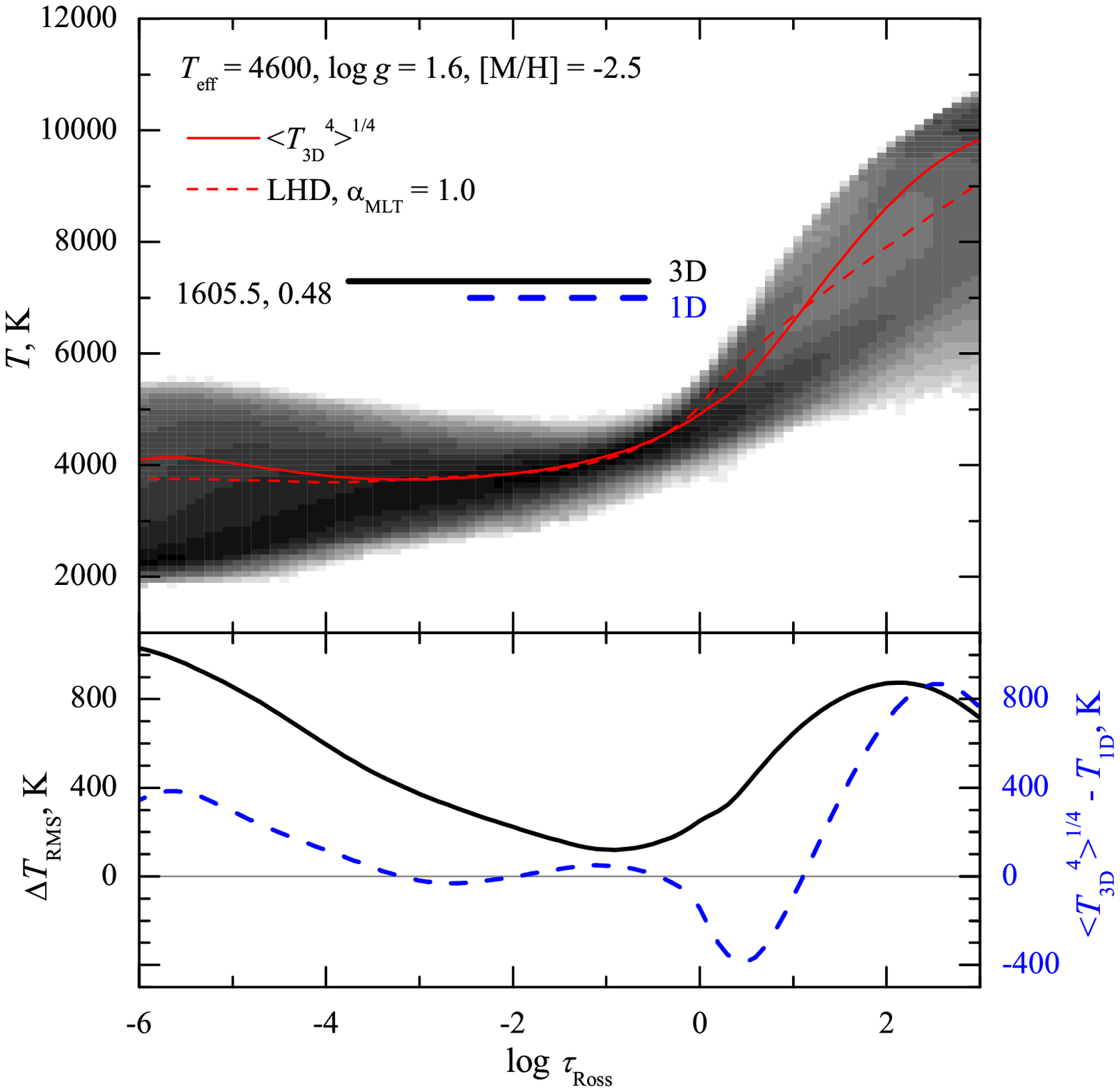}
\caption{\textbf{Top panel:} temperature profiles in the 3D hydrodynamical (gray scale is temperature probability density), average $\xtmean{\mbox{3D}}$ (red solid line), and 1D (red dashed line) model atmospheres, plotted versus the logarithm of Rosseland optical depth, $\log \tau_{\rm Ross}$. All models were computed using identical atmospheric parameters ($\teff=4600$~K, $\log g=1.6$, $\MoH=-2.5$), chemical composition, equation of state, and opacities. Horizontal bars indicate the range in optical depth where typical IR~OH line forms (i.e., its where equivalent width increases from 5\% to 95\%; parameters of this particular line were: wavelength $\lambda = 1605.5$~nm, excitation potential $\chi = 0.48$~eV): black and dashed blue bars correspond to the line forming regions in the full 3D and 1D model atmospheres (the equivalent widths of OH lines in both cases were 1.6~pm). \textbf{Bottom panel:} RMS value of horizontal temperature fluctuations in the 3D model (black line) and temperature difference between the average $\xtmean{\mbox{3D}}$ and 1D models (dashed blue line).\label{fig:T-profiles}}
\end{figure}

\section{Results and conclusions}

Results provided in Table~\ref{table:stars-abundances} show that oxygen abundances determined using 3D hydrodynamical model atmospheres are always smaller than those obtained with 1D hydrostatic models: 3D--1D differences for the three stars are in the range of $-0.23 \dots -0.26$~dex. These differences are a result of large horizontal fluctuations of thermodynamic and hydrodynamical quantities in 3D hydrodynamical model atmospheres of these stars that arise because of convective motions in their outer atmospheric layers. Horizontal temperature fluctuations are clearly seen in Fig.~\ref{fig:T-profiles}, where we show temperature profiles of the 3D hydrodynamical, average $\xtmean{\mbox{3D}}$, and 1D model atmospheres computed using identical atmospheric parameters, chemical composition, equation of state, and opacities. In fact, the presence of large temperature fluctuations (which increase towards the outer atmospheric layers characterized with lower $\log \tau_{\rm Ross}$ values) seems to be a general property of different types of stars and is seen routinely in their 3D hydrodynamical model structures \citep[see, e.g.,][]{LK12,KSL13,DKS13}. Large horizontal temperature fluctuations lead to larger OH line opacities in the low-temperature regions. This, in turn, results in stronger lines in 3D, negative abundance corrections, and thus, lower 3D~LTE abundances.

In fact, it is not only the horizontal temperature fluctuations that influence the formation and strengths of IR~OH lines. The differences between the average temperature profile of the 3D model and that in 1D hydrostatic model do play a role, too. Since the average $\xtmean{\mbox{3D}}$ model is slightly hotter in the optical depths where IR~OH lines form ($\log \tau_{\rm Ross}\approx-0.5\dots-4$; see Fig.~\ref{fig:T-profiles}), this results in somewhat weaker lines in $\xtmean{\mbox{3D}}$, and thus, small but positive abundance corrections. This slightly diminishes the effect of horizontal temperature fluctuations and leads to smaller (i.e., in the absolute sense) total 3D--1D abundance corrections.

At the same time, we obtain a good agreement between the 3D~LTE oxygen abundances and those obtained in 1D~LTE using forbidden [O~I] 630~nm line. The agreement is especially good in HD~122563 and HD~186478, where the difference between abundances obtained using the two indicators is smaller than $\sim0.02$~dex. Although differences are larger in BD~--18:5550, note that in this case 1D~LTE abundance measurement is less reliable, as indicated in \citet[][]{SCP05}.

The obtained results therefore demonstrate that the use of 3D hydrodynamical model atmospheres may allow to reconcile oxygen abundances obtained from IR~OH lines and [O~I] 630~nm line. Nevertheless, one should note that NLTE effects may play a role in the formation of OH lines at low metallicities. The impact here may be two-fold. First, formation and dissociation of OH molecules may be affected by non-equilibrium effects and thus non-equilibrium OH number densities may be different from those where equilibrium molecular formation is assumed. Second, IR~OH lines may form in the conditions that are far from LTE, thus NLTE line strengths may be different from those that would be expected in LTE. It is quite likely that both effects may be important in the atmospheres of EMP red giants, especially in the outer atmospheric layers were molecular lines form (note, however, that the forbidden [O~I] 630~nm line seems to be insensitive to 3D hydrodynamical and NLTE effects, see, e.g., \citealt{A05,DKS13}). In the future, these effects should be properly taken into account using full 3D~NLTE analysis methodology.

\acknowledgments{This work was supported by grants from the Research Council of Lithuania (MIP-065/2013) and the bilateral French-Lithuanian programme ``Gilibert'' (TAP~LZ~06/2013, Research Council of Lithuania). E.C. is grateful to the FONDATION MERAC for funding her fellowship. H.G.L. acknowledges financial support by the Sonderforschungsbereich SFB 881 ``The Milky Way System'' (subproject A4 and A5) of the German Research Foundation (DFG).}

\end{document}